\documentclass[twocolumn,showpacs,preprintnumbers,amsmath,amssymb]{revtex4}
\input psfig.sty

\usepackage{graphicx}
\usepackage{dcolumn}
\usepackage{bm}

\begin{document}

\def\be{\begin{equation}}
\def\ee{\end{equation}}

\title{Age Constraints on Brane Models of Dark Energy}

\author{J. S. Alcaniz} \email{alcaniz@astro.washington.edu}
\affiliation{Astronomy Department, University of Washington, Seattle,
Washington, 98195-1580, USA
}%

\author{D. Jain} \email{deepak@ducos.ernet.in}  \author{A. Dev} \email{abha@ducos.ernet.in} 
\affiliation{%
Department of Physics and Astrophysics, University of Delhi,
Delhi-110007, India  
}%

\date{\today}

\begin{abstract}

Inspired by recent developments in particle physics, the so-called brane world cosmology seems to 
provide an alternative explanation for the present dark energy problem. In this paper, we use the 
estimated age of high-$z$ objects to constrain the value of the cosmological parameters in some 
particular scenarios based on this large scale modification of gravity. We show that such models 
are compatible with these observations for values of the crossover distance between the 4 and 5 
dimensions of the order of $r_c \leq 1.67H_o^{-1}$.

\end{abstract}

\maketitle


The revolutionary ideas associated with extra dimension brane world cosmologies have opened up new
perspectives for a better understanding about the structure of the universe. The general
principle behind such models is that our 4-dimensional Universe 
would be a surface or a brane embedded into a
higher dimensional bulk space-time on which gravity can propagate \cite{ark,dvali,deff} (see also 
\cite{randall} for a recent review). In the recent literature, different aspects of brane world 
cosmologies have been explored. For example, the issue related to the cosmological constant problem 
has been addressed \cite{ccp} as well as cosmological perturbations \cite{brand}, cosmological phase
transitions \cite{cpt}, inflationary 
solutions \cite{cpt1}, baryogenesis \cite{dvali99}, stochastic background of gravitational waves 
\cite{hogan1}, singularity, homogeneity, flatness and entropy problems \cite{aaa},  among others.

An interesting feature of some particular brane world scenarios is the possibility of an accelerated expansion
at the late stages of the cosmic evolution with no need to invoke either a cosmological constant or a
\emph{quintessence} component \cite{deff,deff1}. Such models not only avoid the well known ``cosmological
constant problem"
($\Lambda$ is set to be null) but also enable a description of the presently accelerated stage of the Universe
within the String/M Theory context \cite{Fis,deff1} (see also \cite{sahni} for a discussion on this topic).
From the observational viewpoint, these particular scenarios seem to be in agreement with the most recent
observations of cosmic microwave background and type Ia supernovae \cite{deffZ,dnew} (see, however, 
\cite{avelino}) as well as with the measurements of the angular size of high-$z$ radio sources 
\cite{alcaniz} and the current gravitational lensing data \cite{deepak}.

In this paper we discuss new observational constraints on brane world cosmologies from age considerations due
to the existence of some recently reported old high-redshift galaxies (OHRGs), namely, the LBDS 
53W091, a 3.5-Gyr-old radio galaxy at $z = 1.55$ \cite{dunlop} and the LBDS 53W069, a 4.0-Gyr-old 
radio galaxy at $z = 1.43$ \cite{dunlop1}. We focus our attention on models based on 
the framework of brane-induced gravity of Dvali {\it et al.} \cite{dvali} that have been recently  
proposed in Refs. \cite{deff,deff1}. To be consistent with the inflationary flatness prediction as 
well as with the latest cosmic microwave background (CMB) measurements we restrict ourselves to flat models.
In our analysis we 
assume that the the bulk space-time is 5-dimensional.

The geometry of our 4-dimensional Universe is described by the Friedmann-Robertson-Walker (FRW) 
line element ($c = 1$)
\begin{equation}
ds^2 = dt^2 - R^{2}(t) \left[dr^{2}  + r^{2} (d
\theta^2 + \rm{sin}^{2} \theta d \phi^{2})\right],
\end{equation}
where $r$, $\theta$, and $\phi$ 
are dimensionless comoving coordinates and $R(t)$ is the scale factor. The Friedmann's equation 
for the kind of models we are considering is \cite{deff1,deffZ}
\begin{equation} 
{\dot{R(t)}} = R(t)\left[\sqrt{\frac{\rho}{3M_{pl}^{2}} + \frac{1}{4r_{c}^{2}}} + 
\frac{1}{2r_{c}}\right],
\end{equation} 
where $\rho$ is the energy density of the cosmic fluid, $r_c = M_{pl}^{2}/2M_{5}^{3}$ is the 
crossover scale defining the gravitational interaction among particles located on the brane and  
$M_{pl}$ and $M_5$ are, respectively, the 4- and 5-dimensional Planck mass. As explained in 
\cite{deff}, for distances smaller than $r_c$ the force experienced 
by two punctual sources is the 
usual 4-dimensional gravitational $1/r^{2}$ force whereas for distances larger than $r_c$ the 
gravitational force follows the 5-dimensional $1/r^{3}$ behavior.

Througout this paper we will consider only one component of nonrelativistic particles together with the
bulk-induced term. In this case, the age-redshift relation as a function of the observable parameters is
written as 
\begin{eqnarray}  
t_z  =  {1 \over H_o} & & \int_{0}^{(1 + z)^{-1}} {dx \over x f(\Omega_{\rm{m}}, \Omega_{\rm{r_c}}, 
x)} \\ \nonumber & & = {1 \over H_o} g(\Omega_{\rm{m}}, \Omega_{\rm{r_c}}, z).
\end{eqnarray} 
In the above expression, $x$ is a convenient integration variable and the dimensionless function 
$f(\Omega_{\rm{m}}, 
\Omega_{\rm{r_c}}, x)$ is given by
\begin{eqnarray}
f(\Omega_{\rm{m}}, \Omega_{\rm{r_c}}, x)  = \sqrt{\Omega_{\rm{r_c}}} + 
\sqrt{\Omega_{\rm{r_c}} 
+ \Omega_{\rm{m}}x^{-3}},  
\end{eqnarray}
where $\Omega_{\rm{m}}$ is the matter density parameter and 
\begin{equation}
\Omega_{\rm{r_c}} = 1/4r_c^{2}H_o^{2},
\end{equation} 
is the density parameter associated with the crossover radius $r_c$. From Eq. (2), it is possible to 
show that the flat condition implies
\begin{equation}
\Omega_{\rm{r_c}} = \left(1 - \Omega_{\rm{m}} \over 2 \right)^{2}.
\end{equation}
The total expanding age of the 
Universe is obtained by taking $z = 0$ in Eq. (3). As one may check, 
in the limit $1/r_c \rightarrow 0$, the standard relation [$t_z = \frac{2}{3}H_o^{-1}(1 + 
z)^{-\frac{3}{2}}$] is recovered.

\begin{figure}
\centerline{\psfig{figure=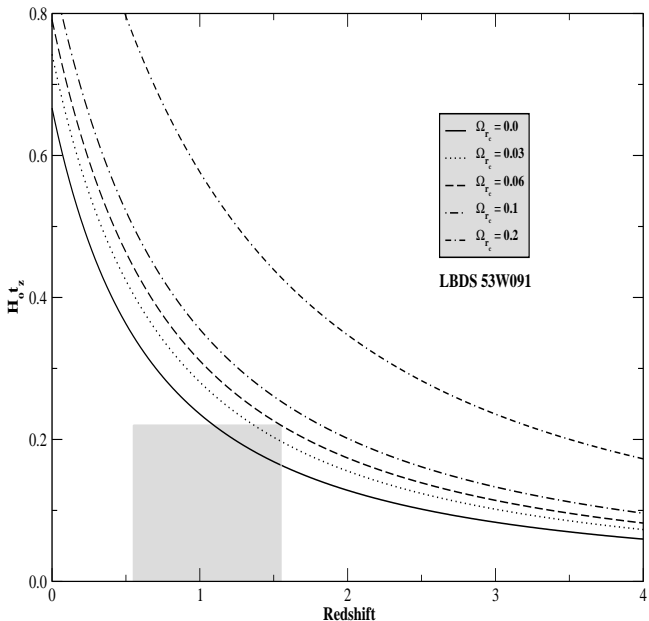,width=3.5truein,height=3.5truein,angle=-90} 
\hskip 0.1in} 
\caption{Dimensionless age parameter as a function of redshift for brane world comologies. As 
explained in the text, all curves crossing the shadowed area yield an age parameter smaller than 
the minimal value required by the galaxy LBDS 53W091 reported in Ref. \cite{dunlop}.}
\end{figure}

We observe that by fixing the product $H_ot_z$ from observations, the limits on the  
parameter $\Omega_{\rm{r_c}}$ can be readily obtained from Eq. (3). Note 
also that the age parameter depends only on the product of the two quantities $H_o$ and 
$t_z$, which are measured from completely different methods. To clarify these points, in the 
following we briefly outline our main assumptions for this analysis. Our approach is based on Ref. 
\cite{jailson}.

\begin{figure}
\centerline{\psfig{figure=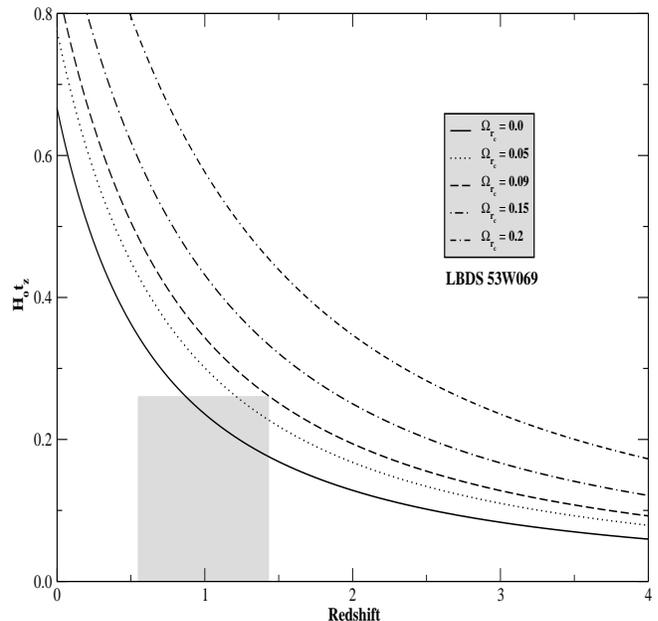,width=3.5truein,height=3.5truein,angle=-90} 
\hskip 0.1in} 
\caption{The same as in Figure 1 for the LBDS 53W069.}
\end{figure}

Firstly, we take for granted that the age of the Universe at a given redshift is bigger than or at 
least equal to the age of its oldest objects. As one may conclude, in the brane world scenarios 
discussed here, the comparison of these two quantities implies a lower bound for 
$\Omega_{\rm{r_c}}$, since the age of the Universe increases for larger values of 
this quantity. As is well known, for the dark and barionic components, the age of the 
Universe increases 
when $\Omega_{\rm{m}}$ decreases, thereby implying the existence of an upper bound for the matter 
densty parameter \cite{kolb}. In order to quantify these qualitative arguments, it is convenient to 
introduce 
the ratio
\begin{equation}
\frac{t_z}{t_g} = \frac{g(\Omega_{\rm{m}}, \Omega_{\rm{r_c}}, z)}{H_o t_g} \geq 1,
\end{equation}
where $t_g$ is the age of an arbitrary object, say, a galaxy at a given redshift $z$ and 
$g(\Omega_{\rm{m}}, \Omega_{\rm{r_c}}, z)$ is the dimensionless factor defined in Eq. 
(3). For each extragalatic object, the denominator of the above equation defines a dimensionless 
age parameter $T_g = H_o t_g$. In particular, the 3.5-Gyr-old galaxy (53W091) at $z = 1.55$ yields 
$T_g = 3.5H_o$Gyr which, for the most recent determinations of the Hubble parameter, $H_o = 70 \pm 
7$ ${\rm{km s^{-1} Mpc^{-1}}}$ \cite{friedman}, takes values on the interval $0.22 \leq T_g \leq 
0.27$. It thus follows that $T_g \geq 0.22$. Therefore, for a given value of $H_o$, only models 
having an expanding age bigger than this value at $z = 1.55$ will be compatible with the existence 
of this object. In particular, taking $1/r_c \rightarrow 0$ (the standard flat case) in Eq. (3), one 
obtains $T_g \leq 0.16$, which means that the Einstein-de Sitter model is formally ruled out by 
this 
test \cite{jailson}.

In order to assure the robustness of the limits, two conditions have been systematically adopted in 
our computations, namely, the minimal value of the Hubble parameter, $H_o = 63$ ${\rm{km s^{-1} 
Mpc^{-1}}}$, and the underestimated age of the galaxies. Such conditions are almost 
self-explanatory when interpreted in the spirit of inequality (7). First, because the smaller the  
value of $H_{o}$, the larger the age that is predicted by the model, and, second, because objects 
with smaller ages are more easily accommodated by the cosmological models, thereby guaranteeing 
that conservative bounds are  
always favored in the estimates presented here. Naturally, similar considerations may also be 
applied to the 4.0-Gyr-old galaxy (53W069) at $z = 1.43$. In this case, one obtains $T_g \geq 0.26$.

In Figs. 1 and 2 we show the diagrams of the dimensionless age parameter $H_ot_z$ as a function 
of the redshift for several values of $\Omega_{\rm{r_c}}$ for the LDBS 53W091 and LBDS 53W069, 
respectively. The forbidden regions in the graphs have been determined by taking the values of 
$T_g$ for each galaxy separately. All curves crossing the shadowed rectangle yield an age parameter 
smaller than the minimal value required by these objects. From this analysis we see that the curves 
intersecting the right upper corner of the rectangle correspond to $\Omega_{r_{c}} = 0.06$ (53W091) and 
$\Omega_{r_{c}} = 0.09$ (53W069). These values establish the lower limites on $\Omega_{r_{c}}$ 
allowed by these two galaxies and provide a minimal total age of the Universe of 12.3 and 13.1 Gyr, 
respectively. Substituting these values of $\Omega_{r_{c}}$ 
into Eq. (5), it is possible to estimate the crossover distance between the 4-dimensional and 
5-dimensional gravities. In this case, for the LBDS 53W091 and LBDS 53W069 bounds, we obtain, 
respectively,
\begin{subequations} 
\begin{equation}
r_c \leq 2.04H_o^{-1}
\end{equation}
and
\begin{equation}
r_c \leq 1.67H_o^{-1}.
\end{equation}
\end{subequations}

\begin{table}
\caption{Recent estimates of the crossover radius $r_c$}
\begin{ruledtabular}
\begin{tabular}{lcr}
Method&Reference&$r_{c}$\footnote{In units of $H_o^{-1}.$}\\
\hline \hline \\
SNe Ia + CMB& \cite{deffZ} & $1.4$\\
SNe Ia & \cite{avelino} & $1.4$\\
Angular size & \cite{alcaniz} & $0.94$\\
Gravitational Lenses & \cite{deepak} & $1.76$\\
OHRGs: & & \\
 $z = 1.55$.........& This paper & $\leq 2.04$\\
 $z = 1.43$.........& This paper & $\leq 1.67$\\
\end{tabular}
\end{ruledtabular}
\end{table}

At this point, it is interesting to compare our estimates of the crossover radius $r_c$ with some 
other recent determinations of this quantity from independent methods. Recently, Deffayet {\it et 
al.} \cite{deffZ} using type Ia supernovae (SNe Ia) + cosmic microwave background (CMB) data found 
$r_c \simeq 1.4H_o^{-1}$ for a flat model with $\Omega_{\rm{m}} = 0.3$. Avelino and Martins 
\cite{avelino}, using a large sample of 92 supernova obtained an approximate relation for the degeneracy in
the $\Omega_{r_{c}} - \Omega_{\rm{m}}$ plane given by $\Omega_{r_{c}} \simeq 
\frac{2}{5}\Omega_{\rm{m}} + \frac{1}{10}$. This approximate best fit line intersects the flat universe curve
in such a way that we can use Eq. (6) to
find
$\Omega_{r_{c}} \simeq 0.12$ ($\Omega_{\rm{m}} = 0.3$) or, equivalently, $r_c \simeq 1.4H_o^{-1}$. More
recently, it 
was shown that measurements of the angular size of high-$z$ sources requires a slightly closed 
universe with a crossover radius of the order of $\sim 0.94H_o^{-1}$ \cite{alcaniz} whereas the 
current gravitational lensing data implies $r_c = 1.76H_o^{-1}$ \cite{deepak}. These results, 
together with the main estimates of the present paper, are summarized in Table 1.

We have investigated new observational constraints for some particular scenarios of brane world 
cosmology. Such models, inspired by recent developments in particle physics, constitute an 
interesting arena in which cosmological tests can play the important role of establishing a 
more solid conection between fundamental physics and astronomical observations. We have shown that 
these scenarios are compatible with the most recent observations of old galaxies at high redshifts for values
of the crossover distance $r_c \leq 1.67H_o^{-1}$. To assure the robustness of the limits derived here we have 
adopted in our computations the minimal value of the Hubble parameter given by \cite{friedman}, i.e., $H_o =
63$ ${\rm{km s^{-1}Mpc^{-1}}}$, as well as the underestimated age of the galaxies. Our results indicate that,
similarly to models with a relic cosmological constant, there is no ``age of the Universe problem" in the
context of these brane world scenarios. Naturally, only with a new and more precise set of observations will
be
possible to show whether or not this class of Superstring-M inspired models constitutes a viable possibility
for the present dark energy problem.
  
\begin{acknowledgments}
The authors are grateful to R. Silva for helpful discussions and a critical
reading of the manuscript. JSA is supported by the Conselho Nacional de Desenvolvimento Cient\'{\i}fico e 
Tecnol\'{o}gico (CNPq - Brasil) and CNPq (62.0053/01-1-PADCT III/Milenio).
\end{acknowledgments}


\end{document}